\documentclass[aps,prd,twocolumn,titlepage,superscriptaddress,nofootinbib]{revtex4-2}

\usepackage{graphicx}      
\usepackage{amssymb}       
\usepackage{amsmath}       
\usepackage{latexsym}      
\usepackage{cancel}        
\usepackage[normalem]{ulem} 
\usepackage{url}           
\usepackage{soul}          
\usepackage{verbatim}      
\usepackage{multirow}      
\usepackage{mathrsfs}      
\usepackage{float}         
\usepackage[dvipsnames]{xcolor} 
\usepackage{mathtools}     
\usepackage{slashed}       
\usepackage{physics}       
\usepackage{epstopdf}      
\usepackage{subfigure}     
\usepackage{bbold}         
\usepackage{wasysym}       
\usepackage{feynmp}        
\usepackage[colorlinks=true, citecolor=blue, linkcolor=blue, urlcolor=blue]{hyperref}
\usepackage{orcidlink}
\usepackage{etoolbox}

\newcommand{\beq}{\begin{eqnarray}}
\newcommand{\eeq}{\end{eqnarray}}

\makeatletter
\newcommand{\myBig}{\bBigg@{1.75}}
\makeatother

\begin{document}
\title{ Nonequilibrium crossover in the supercritical region from quench dynamics}

\author{Zi-Qiang Zhao\orcidlink{0009-0009-7859-3655}}
\affiliation{Liaoning Key Laboratory of Cosmology and Astrophysics, College of Sciences, Northeastern University, Shenyang 110819, China}
\author{Zhang-Yu Nie\orcidlink{0000-0001-7064-247X}}\email{niezy@kust.edu.cn}
\affiliation{Center for Gravitation and Astrophysics, Kunming University of Science and Technology, Kunming 650500, China}
\author{Jing-Fei Zhang\orcidlink{0000-0002-3512-2804}}
\affiliation{Liaoning Key Laboratory of Cosmology and Astrophysics, College of Sciences, Northeastern University, Shenyang 110819, China}
\author{Xin Zhang\orcidlink{0000-0002-6029-1933}}\email{zhangxin@neu.edu.cn}
\affiliation{Liaoning Key Laboratory of Cosmology and Astrophysics, College of Sciences, Northeastern University, Shenyang 110819, China}
\affiliation{MOE Key Laboratory of Data Analytics and Optimization for Smart Industry, Northeastern University, Shenyang 110819, China}
\affiliation{National Frontiers Science Center for Industrial Intelligence and Systems Optimization, Northeastern University, Shenyang 110819, China}

\begin{abstract}
Distinguishing different subphases in the supercritical region is an important issue in statistical physics and condensed matter physics. Traditional approaches rely mainly on static thermodynamic response functions or equilibrium correlation functions, which are essentially limited to quasistatic processes. In this paper, we investigate the evolution behavior of a system after a rapid quench from the perspective of nonequilibrium dynamics within a holographic model. We find that, using the time at which inhomogeneous structures appear most rapidly, we can define a supercritical crossover curve based on the pure phase separation process. In addition, the uniform invasion phenomenon induced by topological defects persists in the supercritical region, and the invasion velocity exhibits a clear turning point as a function of the quench endpoint. This turning point can define another new nonequilibrium supercritical crossover line that simultaneously incorporates the effects of both symmetry breaking and phase separation. Unlike the classical Widom line or Frenkel line, these two new crossover lines contain both thermodynamic information and dynamical information, reflecting the dynamical nature of the supercritical region under nonequilibrium conditions. This work provides a novel nonequilibrium dynamical approach for characterizing supercritical subphases.
\end{abstract}
\maketitle
\section{Introduction}
The critical phenomenon is a ubiquitous phenomenon that appears in quantum systems \cite{PhysRev.149.118,Son:2020yiy,Jim_nez_2021}, classical systems (e.g., water \cite{10.1063/PT.3.1796}), and strongly gravitating systems \cite{Kubiznak:2012wp,Kubiznak:2016qmn,Cai:2013qga,Wei:2014hba}. Beyond the critical point, the system enters the supercritical region, where the first-order phase transition disappears and the two distinct phases can no longer be distinguished by conventional methods. However, recent research shows that one can still identify novel supercritical subphases using thermodynamic quantities (e.g., Widom line\cite{Xu_2005,Ruppeiner_2012,PhysRevLett.112.135701,PhysRevE.95.052120,Gallo2014}) or dynamical probes (e.g., Frenkel line \cite{Yoon_2018,PhysRevLett.111.145901,Bolmatov2013,Prescher_2017,Bolmatov_2015,Fomin_2018,Fomin2015,PhysRevE.85.031203,2023PhRvR...5a3149H,jiang2024experimental}). Moreover, in black hole systems with first-order phase transitions, different subphases can also be defined via Lee-Yang phase transition theory, quasinormal modes, or other various methods \cite{Zhao:2024jhs,Xu:2025jrk,Zhao:2025ecg,Wang:2025ctk,Li:2025lrq,Guo:2026xlk,Zhao:2025ecg,Li:2025tdd,Anand:2026pfz}.

Nevertheless, all the aforementioned approaches are based on quasi-static or equilibrium conditions. In nature, however, evolutions often occur on very short timescales, implying that purely quasi-static results may not fully capture all the information in a rapid evolution process. Furthermore, the study of the supercritical region in holographic systems inevitably involves the dynamic evolution of symmetry breaking. Traditional methods only consider information near the critical point of first-order phase transition, so information on symmetry breaking is likely missing. Therefore, exploring the properties of the supercritical region through nonequilibrium dynamics via a fast quench across the critical point is a highly reasonable and necessary approach.

The holographic model constructed via the AdS/CFT duality \cite{Maldacena:1997re} provides a highly powerful framework for investigating the nonequilibrium dynamical evolution of such strongly coupled systems \cite{Hartnoll:2008vx,Hartnoll:2008kx,Herzog:2010vz,Chen:2022cwi,Li:2020ayr,Chen:2022tfy,Xia:2020cjl,delCampo:2021rak,Li:2021iph,Li:2021dwp,Xia:2021xap,Zeng:2022hut,Yang:2025bsw,Xia:2026yrj,Xia:2023pom,Su:2023vqa,Zhao:2023ffs,An:2024ebg,Yang:2024hom,Xia:2024ton,Zeng:2024rwn}. Moreover, holographic models can also be employed to probe the interior structure of black holes \cite{Hartnoll:2020fhc,Cai:2020wrp,Wang:2020nkd,Liu:2021hap,An:2021plu,Mansoori:2021wxf,Liu:2021hap,Cai:2021obq,An:2022lvo,Sword:2022oyg,Liu:2022rsy,Zhao:2022jvs,Xu:2023fad,Gao:2023zbd,Gao:2023rqc,Zhang:2025hkb,Zhang:2025tsa,Xu:2025edz,Gao:2025qqa,Xiong:2026npi,Zhao:2026mkx}. Compared with the conventional Ginzburg–Landau theory, a notable advantage of the holographic approach lies in its capacity to naturally realize diverse nonequilibrium physical processes, such as phase separation \cite{Janik:2015iry,Janik:2016btb,Janik:2017ykj,Attems:2017ezz,Attems:2019yqn,Bellantuono:2019wbn,Attems:2020qkg,Chen:2022tfy,Ning:2023edr,Zhao:2023ffs,Jin:2026lzp,Zhao:2026als}. When the system is quenched across a critical point, spontaneous symmetry breaking becomes relevant. Holographic models further facilitate the investigation of the combined effects arising from these two mechanisms \cite{Zhao:2026eav}.

According to previous studies in holographic systems \cite{Zhao:2023ffs}, phase separation persists even within the supercritical region, and this supercritical phase separation has been experimentally observed \cite{Lee2025}. This is of considerable importance, as phase separation is theoretically expected to occur exclusively in the first-order phase transition regime. Consequently, investigating the supercritical region via the information from supercritical phase separation constitutes a significant research direction; yet such investigations remain substantially incomplete. In Ref.~\cite{Zhao:2026eav}, the authors demonstrated that when symmetry breaking is present and the quench endpoint resides within the unstable region of the first-order phase transition, topological defects can serve as nucleation sites for phase separation, thereby inducing a uniform invasion phenomenon. A natural question thus arises: can such nonequilibrium phenomenon be exploited to probe the properties of the supercritical region?

Based on the above considerations, in this work we have investigated the pure phase separation phenomenon and the invasion velocity, as well as their dependence on the quench endpoint  $\rho_f$, and defined two new supercritical crossover lines based on nonequilibrium processes. In contrast to conventional approaches, our method is entirely grounded in nonequilibrium dynamics, implying that the supercritical crossover line we define incorporates both thermodynamic and dynamic information. The remainder of this paper is organized as follows. In Section \ref{sec2}, we briefly introduce our model. In Section \ref{sec3}, we present the static solutions and critical phenomena. In Section \ref{sec4}, we discuss the nonequilibrium process and supercritical crossover. Finally, in Section \ref{sec5}, we provide a summary.


\section{The holographic model setup}\label{sec2}
The model we employ is a holographic system, specifically an Einstein–Maxwell–scalar theory with a neutral scalar field that respects a $\mathbb{Z}_2$ symmetry. We introduce two higher‑order nonlinear terms, primarily to induce a first‑order phase transition and to realize supercritical phenomenon. This work will focus on the probe limit, which means that we neglect the backreaction of the matter fields on the spacetime background. In this work, the Lagrangian takes the following form
\begin{align}
\mathcal{L}_{m}=&-\frac{1}{4}h(\Psi)F_{\mu\nu}F^{\mu\nu}-\nabla_{\mu}\Psi \nabla^{\mu}\Psi-m^{2}\Psi^2\nonumber\\
&-\lambda\Psi^4-\tau\Psi^6,\label{Lag}
\end{align}
in which $h(\Psi)=e^{\alpha \Psi^2}$ and $\alpha$ is the coupling parameter. We adopt an ansatz of the form
\begin{align}
\Psi&=z\psi(t,z,x)/L~,\\
A_\mu dx^\mu&=A_t(t,z,x)dt+A_x(t,z,x)dx~.
\end{align}
Since this work involves non‑equilibrium dynamical evolution, it is convenient to adopt the in-going Eddington metric. Accordingly, the metric takes the following form
\begin{align}
ds^{2}=\frac{L^2}{z^2}(-f(z)dt^{2}-2dtdz+dx^{2}+dy^{2})~,
\end{align}
in which $f(z)=1-(z/z_h)^3$ where $z_h$ is the horizon of black hole. The complete dynamical evolution equations are
\begin{widetext}
\begin{align}
    2 \lambda  \psi ^3-\alpha  \left(\partial _tA_x\right) \left(\partial _zA_x\right) \psi  z^2 e^{\alpha  \psi ^2 z^2}+2 \partial _t\partial _z\psi +\alpha  \left(\partial _xA_t\right) \left(\partial _zA_x\right) \psi  z^2 e^{\alpha  \psi ^2
        z^2}-\partial _x\partial _x\psi 
    -\frac{1}{2} \alpha  \left(\partial _zA_t)^2\right. \psi  z^2 e^{\alpha  \psi ^2 z^2}\nonumber\\
    +\frac{1}{2} \alpha  \left(\partial _zA_x)^2\right. \psi  z^2 e^{\alpha  \psi ^2 z^2}-\frac{1}{2} \alpha  \left(\partial
    _zA_x)^2\right. \psi  z^5 e^{\alpha  \psi ^2 z^2}
    +\partial _z\partial _z\psi  z^3-\partial _z\partial _z\psi +3 \left(\partial _z\text{$\psi $)}\right. z^2+3 \tau  \psi ^5 z^2+\psi  z=0&,\label{A1}\\
    -2 \alpha  \left(\partial _x\text{$\psi $)}\right. \left(\partial _zA_x\right) \psi  z^2+2 \alpha  \left(\partial
    _zA_t\right) \psi  z (\left(\partial _z\text{$\psi $)}\right. z+\psi )-\partial _z\partial _xA_x+\partial
    _z\partial _zA_t=0&,\label{A2}\\
    2 \alpha  \left(\partial _tA_x\right) \left(\partial _x\text{$\psi $)}\right. \psi  z^2+\partial _t\partial
    _xA_x+\partial _t\partial _zA_t+2 \alpha  \left(\partial _t\text{$\psi $)}\right. \left(\partial _zA_t\right) \psi 
    z^2-2 \alpha  \left(\partial _xA_t\right) \left(\partial _x\text{$\psi $)}\right. \psi  z^2
    -\partial _x\partial
    _xA_t\nonumber\\
    +2 \alpha  \left(\partial _x\text{$\psi $)}\right. \left(\partial _zA_x\right) \psi  z^5-2 \alpha 
    \left(\partial _x\text{$\psi $)}\right. \left(\partial _zA_x\right) \psi  z^2+\partial _z\partial _xA_x
    z^3-\partial _z\partial _xA_x=0&,\label{A3}\\
    -2 \alpha  \left(\partial _tA_x\right) \left(\partial _z\text{$\psi $)}\right. \psi  z^2-2 \alpha  \left(\partial
    _tA_x\right) \psi ^2 z-2 \partial _t\partial _zA_x
    -\left(\partial _zA_x\right) z \left(-2 \alpha  \psi ^2+2 \alpha 
    \left(\partial _t\text{$\psi $)}\right. \psi  z+2 \alpha  (\partial _z\text{$\psi $)}\right. \psi 
    \left(z^3-1\right) z\nonumber\\
    +2 \alpha  \psi ^2 z^3+3 z)
    +2 \alpha  \left(\partial _xA_t\right) \psi  z (\left(\partial
    _z\text{$\psi $)}\right. z+\psi )+\partial _z\partial _xA_t-\partial _z\partial _zA_x z^3+\partial _z\partial _zA_x=0&.
    \label{A4}
\end{align}
\end{widetext}
Eq.~(\ref{A3}) is a constraint equation, which in the conformal boundary is
\begin{align}
\partial _t\rho=-\partial _x\partial _xA_t-\partial _z\partial _xA_x.
\end{align}

Numerically, we employ the Chebyshev pseudospectral method in the holographic $z$-direction with 21 grid points. In the spatial direction, we adopt the Fourier spectral method, which imposes periodic boundary conditions in the $x$-direction. For the time direction, we use the fourth-order Runge-Kutta method with a time step of $\delta t = 0.05$.


\section{Static solutions and supercritical phenomenon}\label{sec3}
Before presenting the full dynamical evolution, it is necessary to first introduce the static solutions and how the supercritical region is realized. Since our model incorporates two higher-order nonlinear terms, it allows for various types of phase transitions. The full phase diagram in terms of the coefficients of the higher-order terms has been presented in previous work \cite{Zhao:2026eav}.


For static solutions, we adopt the metric of the following form
\begin{align}
ds^{2}=\frac{L^2}{z^2}(-f(z)dt^{2}+\frac{1}{f(z)}dz^{2}+dx^{2}+dy^{2}),
\end{align}
and the ansatz is as follows
\begin{align}\label{ansatz}
\Psi=z\psi(z)/L~, A_\mu dx^\mu=\phi(z)dt~,
\end{align}
in which, $L$ is the AdS radius.
The complete equations of motion for static solutions can be written in the following form
\begin{align}
\psi(\frac{f'}{f z}-\frac{m^2}{f z^2}+\frac{\alpha  z^2 \phi '^2
   e^{\alpha  \psi ^2 z^2}}{2 f}-\frac{2}{z^2})+\frac{f' \psi '}{f}\nonumber\\
   -\frac{2 \lambda  \psi
   ^3}{f}-\frac{3 \tau  \psi ^5 z^2}{f}+\psi ''=0~,\\
   2 \alpha  \psi  z^2 \psi ' \phi '+2 \alpha  \psi ^2 z \phi '+\phi ''=0~,
\end{align}
the boundary conditions are
\begin{align}
\phi(z)= \mu-z\rho+\dots~,~\psi(z)= \psi^{(1)}+ z\psi^{(2)}+\dots~.
\end{align}
The free energy is
\begin{align}
	G=\frac{\widetilde{G}T}{V_2}=\frac{\mu\rho}{2L^2}+\int_{0}^{z_h}\bigg(\frac{1}{2}e^{z^2\alpha\psi^2}z^2\alpha\psi^2\phi'^2\nonumber\\
    -\lambda\psi^4-2z^2\tau\psi^6\bigg)dz,\label{FEconoical}
\end{align}
where $V_2$ is the volume of the spatial boundary manifold. In this model, the black hole temperature is fixed as $T = 3/(4\pi z_h)$, and the system is characterized by the charge density, implemented via the boundary condition $A_t'(\infty) = -\rho$. We adopt the standard quantization scheme, where $\psi^{(1)}(\infty) = 0$ and the vacuum expectation value is given by $\langle O_2 \rangle= \psi^{(2)}/\rho$. In the remainder of this work, we set $\alpha = 5$, $L = 1$, and $z_h = 1$. Here we present only the free energy expression for the canonical ensemble. For the grand canonical ensemble, the corresponding free energy is obtained by applying a Legendre transformation to Eq.~(\ref{FEconoical})


\begin{figure}
\centering
\includegraphics[width=0.9\columnwidth]{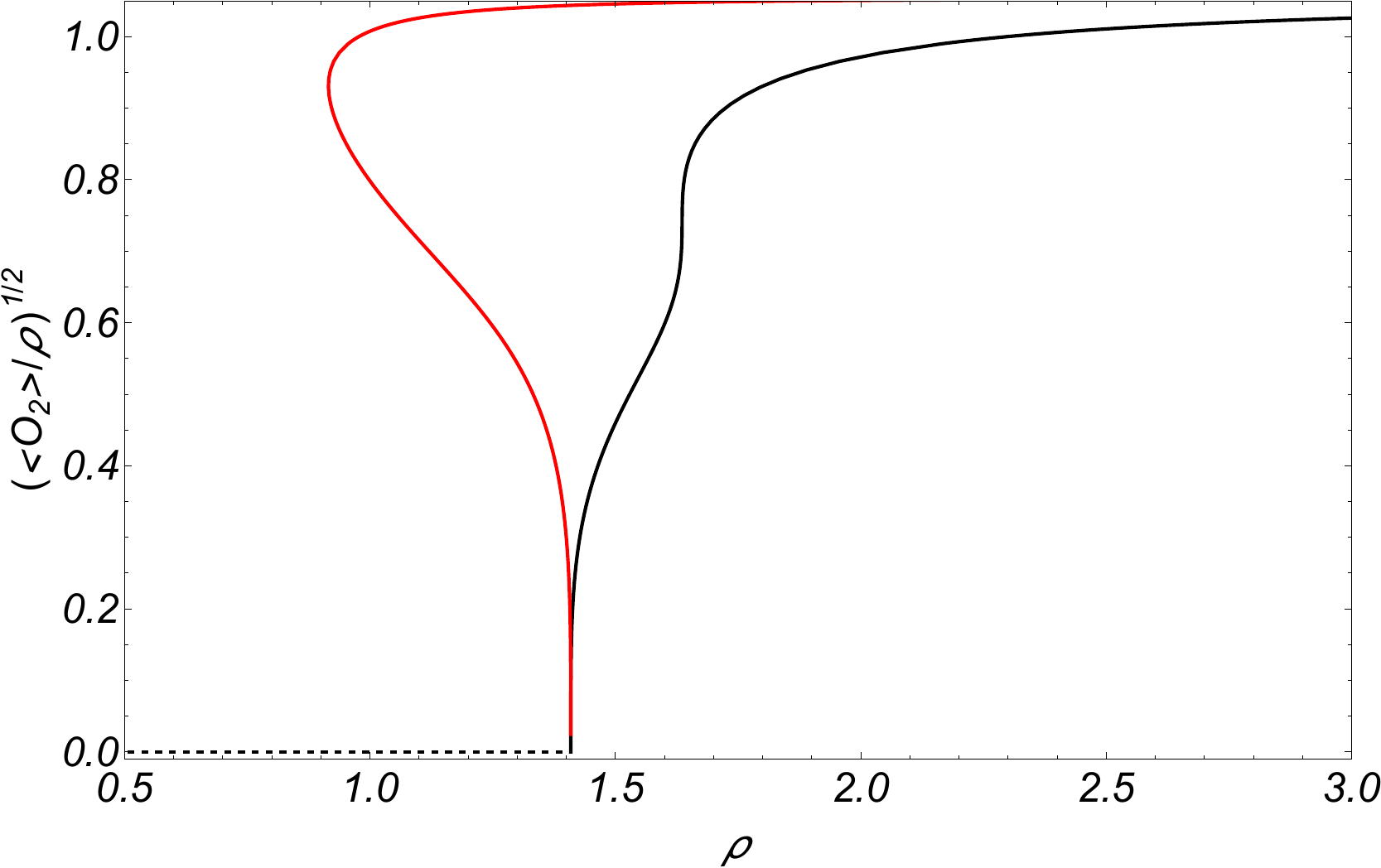}
\includegraphics[width=0.9\columnwidth]{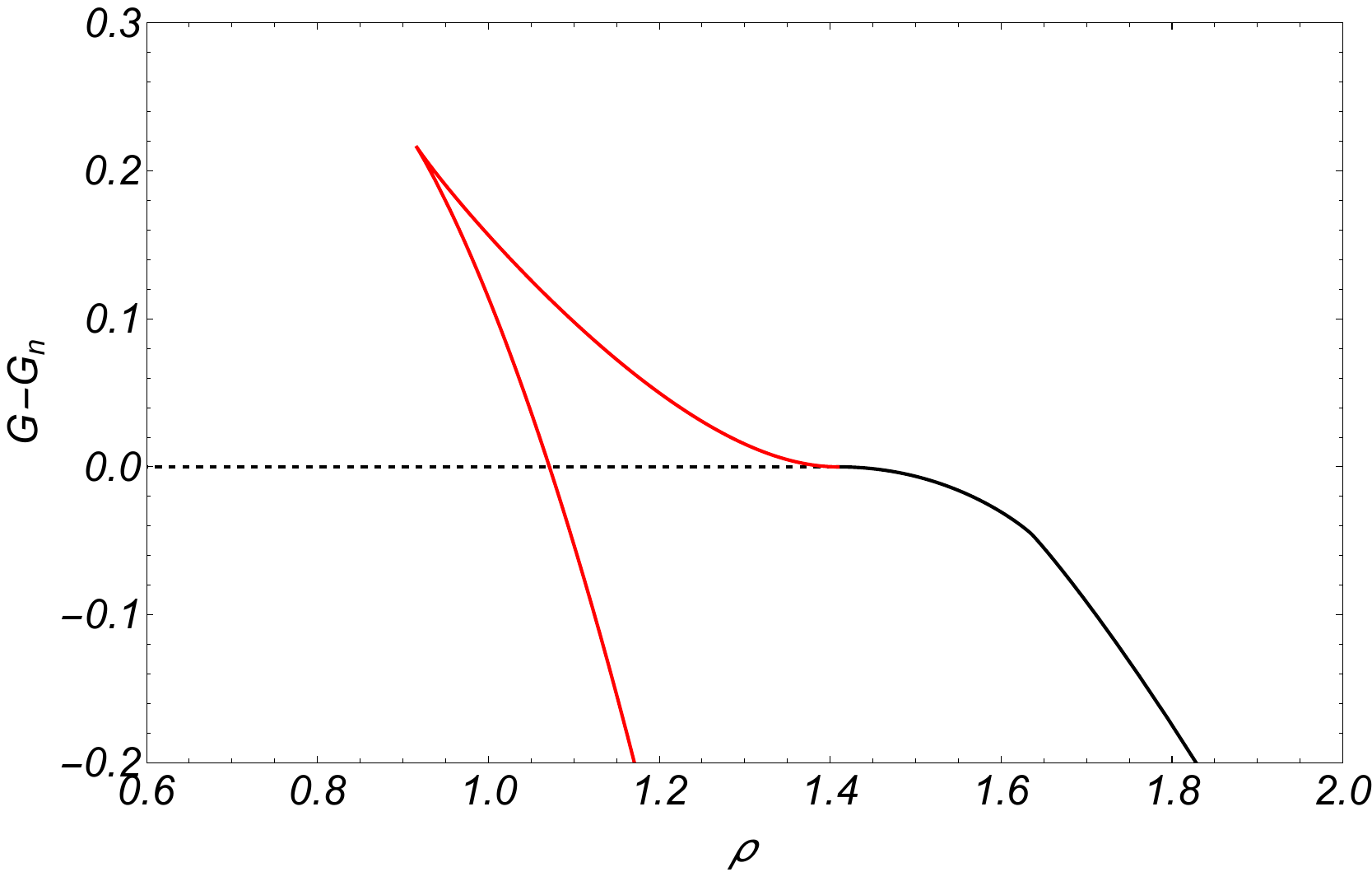}
\caption{Condensate and free energy of the neutral scalar field. The black curves correspond to the canonical ensemble (fixed $\rho$), and the red curves correspond to the grand canonical ensemble. Theoretically, the vertical coordinate for the grand canonical ensemble should be $\mu$, but we plot them together for comparison.}\label{condensate02}
\end{figure}

\begin{figure}
\centering
\includegraphics[width=0.9\columnwidth]{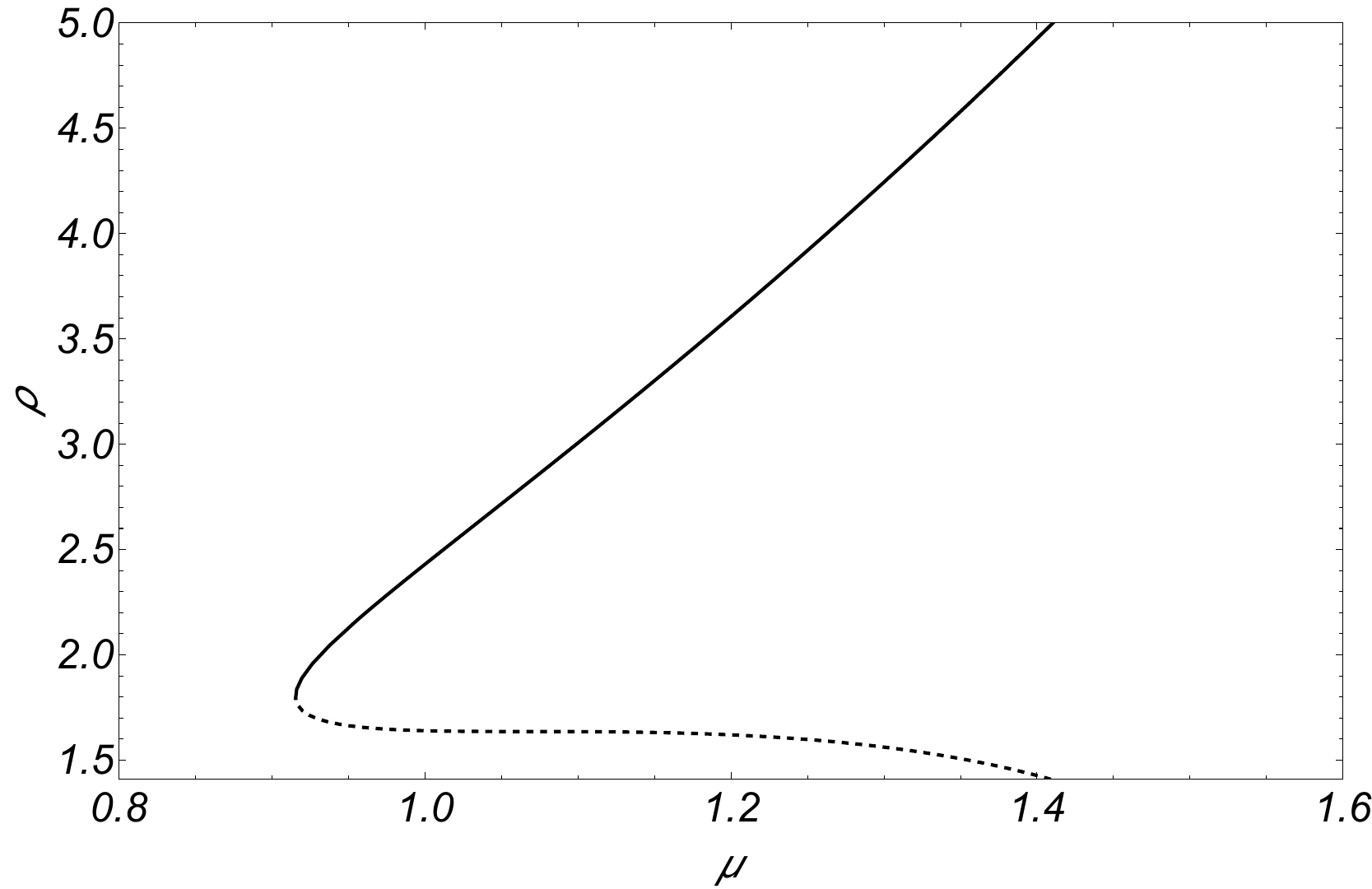}
\caption{The relationship between the charge density $\rho$ and the chemical potential $\mu$. The solid line denotes the stable region, and the dashed line denotes the unstable region, which corresponds to the negative charge susceptibility $\partial \rho/\partial \mu < 0$.}\label{murho}
\end{figure}


First, we choose a negative value for the parameter $\lambda$. In this work, we set $\lambda = -4$ and then turn on the $\tau$ parameter. In principle, as long as $\tau \neq 0$, the system can always find a branch with lower free energy for sufficiently large $\psi$. 
At this point, the critical value of $\tau$ for the system is $\tau_c=2.93$.
If the nonlinear parameter $\tau$ is further increased, the system will enter the supercritical region, where conventional thermodynamic methods are no longer capable of distinguishing the supercritical subphases. Since the first-order phase transition no longer exists, inhomogeneous structures should not be generated during the dynamic quench process in theory. However, in the present system, even though the canonical ensemble enters the supercritical region where inhomogeneous structures are theoretically no longer expected to form, the grand canonical ensemble still possesses an unstable branch, and consequently inhomogeneous structures can still emerge during the quench process. Notably, such supercritical phase separation has already been realized experimentally \cite{Lee2025}.

This process can be understood as follows. The system under consideration is globally charge conserving and thus corresponds to the canonical ensemble. However, during nonequilibrium dynamical evolution, particle exchange can still occur among local subsystems, a scenario described by the grand canonical ensemble, and the dynamical stability of the system coincides with that of the local subsystems. This accounts for the emergence of inhomogeneous structures even after the canonical ensemble has entered the supercritical region (see,  \textit{e.g.}, Refs. \cite{Zhao:2022jvs,Zhao:2023gcv}). In Fig.~\ref{condensate02}, we show the condensate and free energy results for $\lambda=-4$ and $\tau=2.93$. It can be seen that under these parameters, the canonical ensemble has no unstable region, i.e., the system is in the supercritical region, while the grand canonical ensemble still possesses an unstable region. In this case, the interval in which spatially inhomogeneous structures can emerge is equal to the region where $\partial \rho/\partial \mu < 0$ in Fig.~\ref{murho}, i.e., the negative charge susceptibility \cite{Zhao:2022jvs}.


\section{Nonequilibrium quench process and Supercritical crossover}\label{sec4}
Before discussing the supercritical region, we briefly outline the nonequilibrium process considered here. When the system is quenched from the normal solution into the condensate solution across the critical point, symmetry breaking occurs. If the quench endpoint is chosen precisely within the unstable region of the first-order phase transition, phase separation also takes place. Results from Ref.~\cite{Zhao:2026eav} indicate that phase separation occurs after symmetry breaking in the process of fast quench. For random initial perturbations, the system is divided into several separated regions by kinks, so phase separation proceeds almost simultaneously across these regions. However, if a specific initial configuration is imposed (e.g., an initial perturbation with one half positive and the other half negative), only a single pair of kinks remains after the quench. Phase separation then first emerges at the kinks, where spatial inhomogeneity is maximized, and subsequently invades the region between the two kinks at a constant speed, this is called the invasion phenomenon. 

In this paper, we will mainly investigate the supercritical crossover via nonequilibrium quench dynamics. Specifically, we will discuss two different scenarios. First, we consider the case that involves only phase separation, which in principle applies to all systems with a supercritical region. Then we turn to a more complex case where both symmetry breaking and phase separation are present. This scenario may be more applicable to supercritical systems with symmetry breaking, such as the Ising model \cite{Zhao_2021,Orito_2025} or QCD systems \cite{Chen_2021}.


\begin{figure}
\centering
\includegraphics[width=0.9\columnwidth]{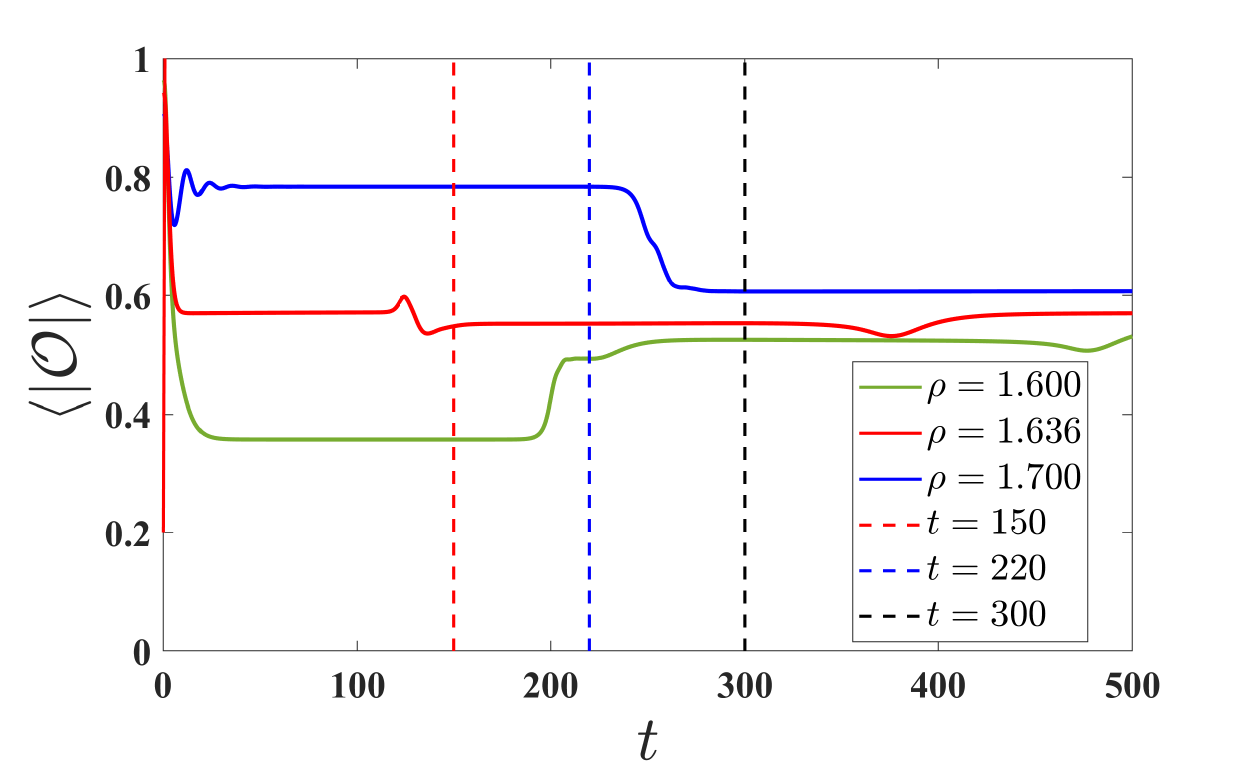}
\includegraphics[width=0.9\columnwidth]{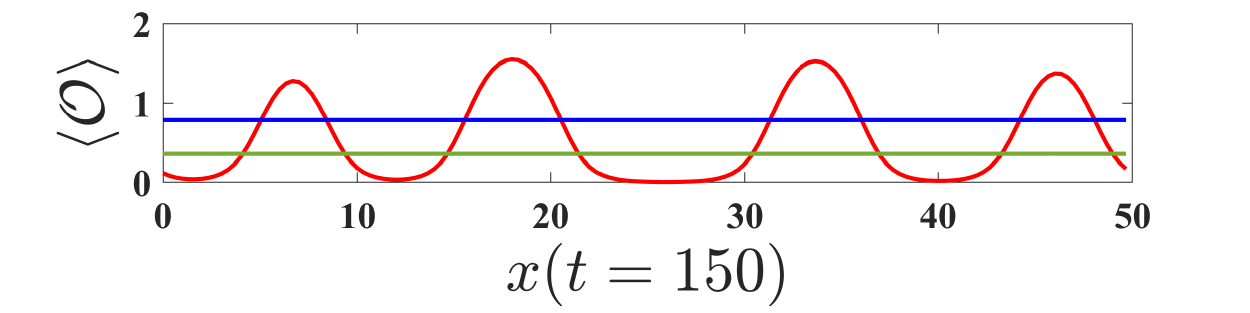}
\includegraphics[width=0.9\columnwidth]{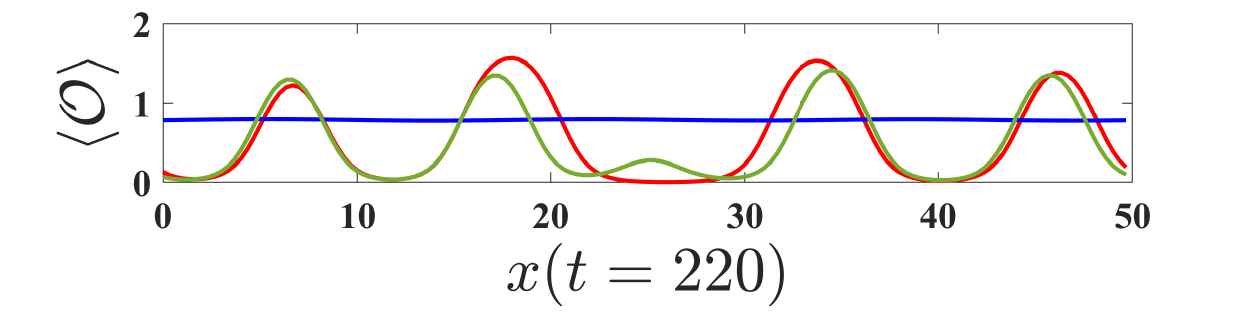}
\includegraphics[width=0.9\columnwidth]{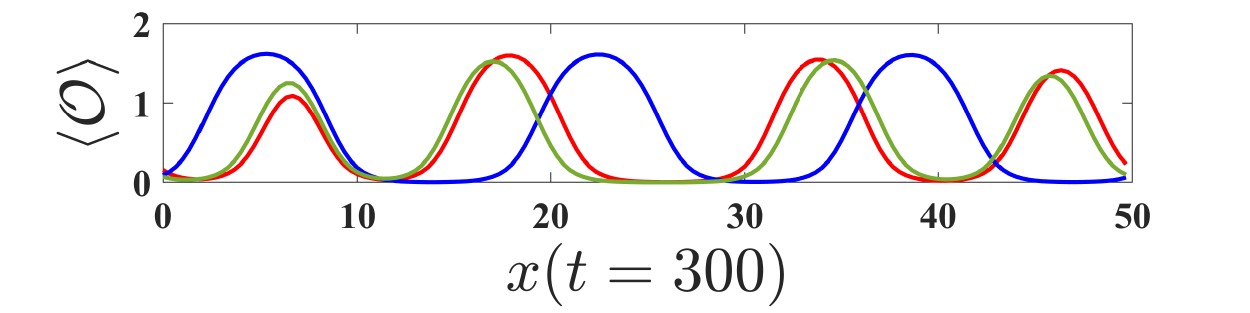}
\caption{The spatial distribution of the system's condensate as a function of time.}\label{diagramxvo}
\end{figure}

\begin{figure}
\centering
\includegraphics[width=0.9\columnwidth]{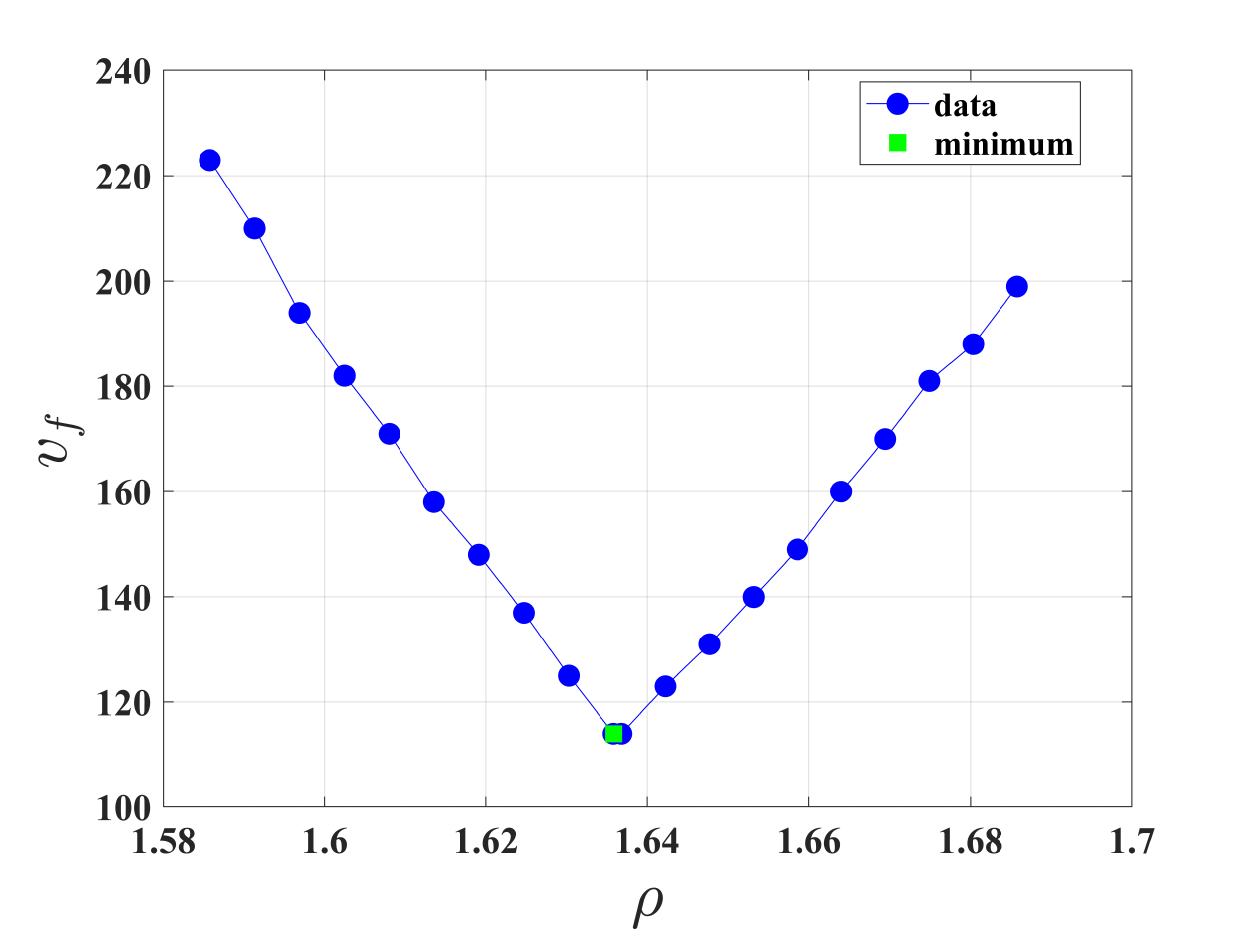}
\caption{The relationship between the time at which spatially inhomogeneous structures emerge most rapidly during phase separation and the quench endpoint $\rho_f$.}\label{examplePhaseSeparation}
\end{figure}

\subsection{Supercritical crossover from pure phase separation}
Let us first discuss the case with only the phase separation effect. In this case, we can choose to start from a uniform condensed state and quench toward the spinodal decomposition region of the system. The system then develops spatially inhomogeneous structures, i.e., bubbles, due to dynamical instability. In our numerical experiment, we set $\rho_i = 1.8$ and quench back at a rate $\tau_Q = 0.1$.  We present the results in Fig.~\ref{diagramxvo}. We find that different $\rho_f$ values correspond to different times for the emergence of inhomogeneous structures. This suggests that there exists a characteristic point in this phase separation effect at which inhomogeneous structures appear most rapidly.


\begin{figure*}[!htbp]
\centering
\includegraphics[width=0.67\columnwidth]{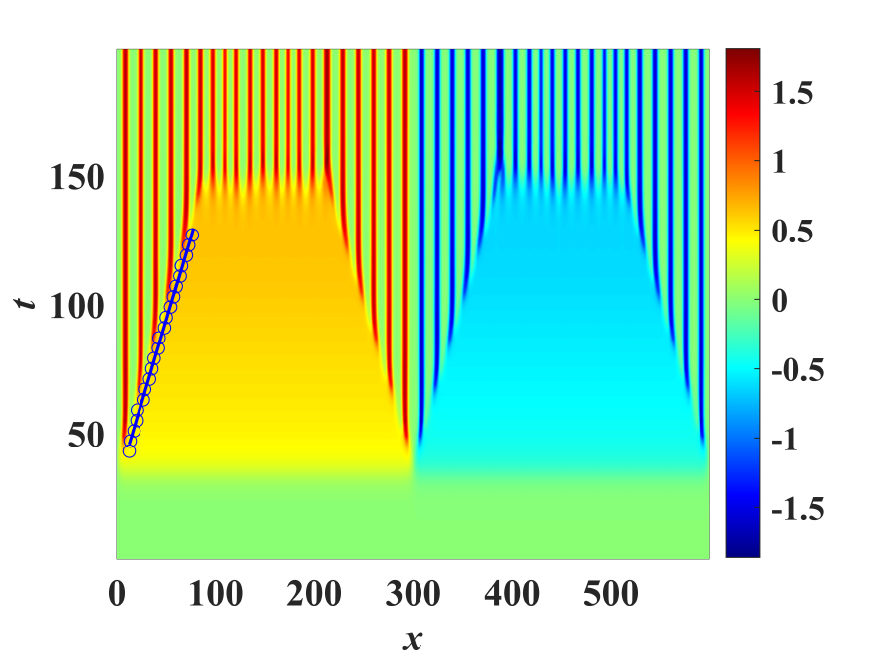}
\includegraphics[width=0.67\columnwidth]{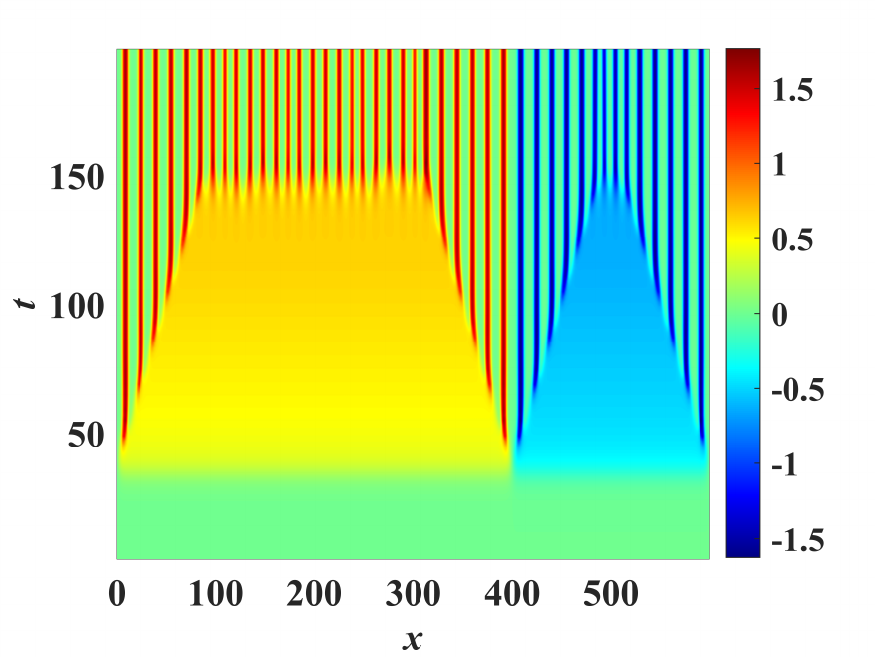}
\includegraphics[width=0.67\columnwidth]{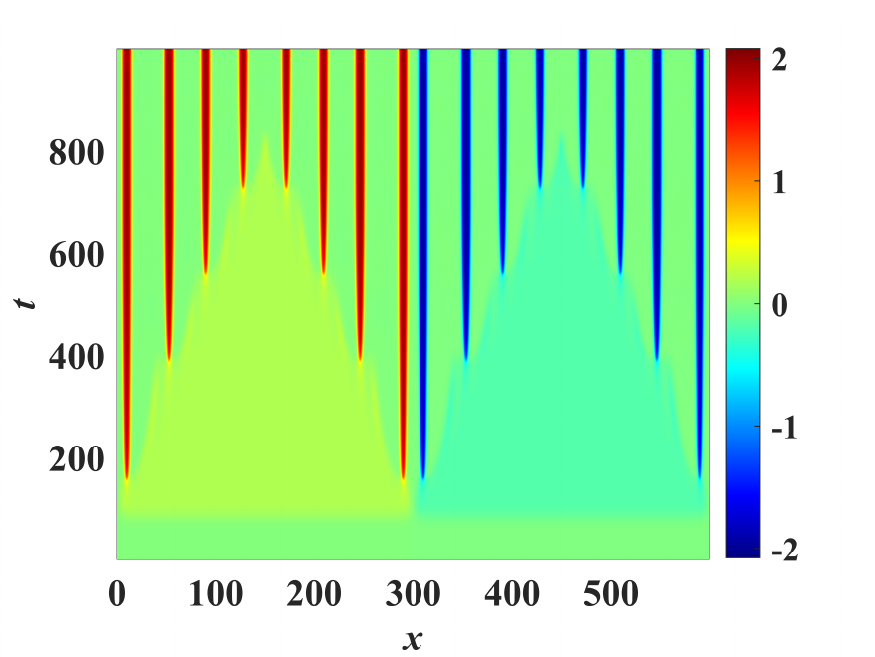}
\caption{The process of invasion phenomenon during a quench in the supercritical region with $L_x=600$, $n_x=1800$, and $\tau_Q=0.1$. Left panel: The invasion phenomenon for quench parameters $\rho_s = 1.3$ and $\rho_f = 1.64$, with the initial condition of $\psi$ given by $\psi_i(x) = \{F_i,0\leq x<L_x/2;-F_i,L_x/2<x\leq L_x\}$ with $F_i=10^{-5}$. Middle panel: The quench parameters are the same as in the left panel, and the initial condition for $\psi$ is set as $\psi_i(x) = \{F_i,0\leq x<L_x*2/3;-F_i,L_x*2/3<x\leq L_x\}$ with $F_i=10^{-5}$. Right panel: The invasion phenomenon for quench parameters $\rho_s = 1.3$ and $\rho_f = 1.48$, and the initial condition is the same as in the left panel. Where the color bar indicates the magnitude of the condensate. The blue circles denote the points where the invasion phenomenon occurs at a fixed time. The blue solid line represents the linear fit to these points, with the fitting function given by $v_{inv}(t) = at + b$. We define $a$ as the invasion velocity $v_i$.}\label{phaseSpinvasionSuper}
\end{figure*}


To verify this, we compute the time at which spatially inhomogeneous structures appear most rapidly for different quench endpoints $\rho_f$, and present the results in Fig.~\ref{examplePhaseSeparation}. It can be seen from the results that, as we speculated earlier, there indeed exists a point at which inhomogeneous structures appear most rapidly. Therefore, for systems with only the pure phase separation effect, we can regard this point as the dynamical supercritical crossover point. In theory, this approach has fewer limitations for studying the supercritical region. As long as the system exhibits supercritical phase separation, this method can be used to determine the supercritical crossover point.


\subsection{Supercritical crossover from invasion phenomenon}
Now let us discuss another scenario. If a system contains both phase separation and symmetry breaking, then determining the supercritical crossover point solely through the pure phase separation effect may miss the information associated with symmetry breaking. For systems with symmetry breaking, such information is very important because this process is often related to intrinsic properties such as the correlation length. Therefore, if this information can be further exploited to investigate the properties of the supercritical region, it would be very helpful for understanding the physical characteristics of the supercritical region. In this paper, we adopt the invasion velocity as our method \cite{Zhao:2026eav}.

\begin{figure}
\centering
\includegraphics[width=0.9\columnwidth]{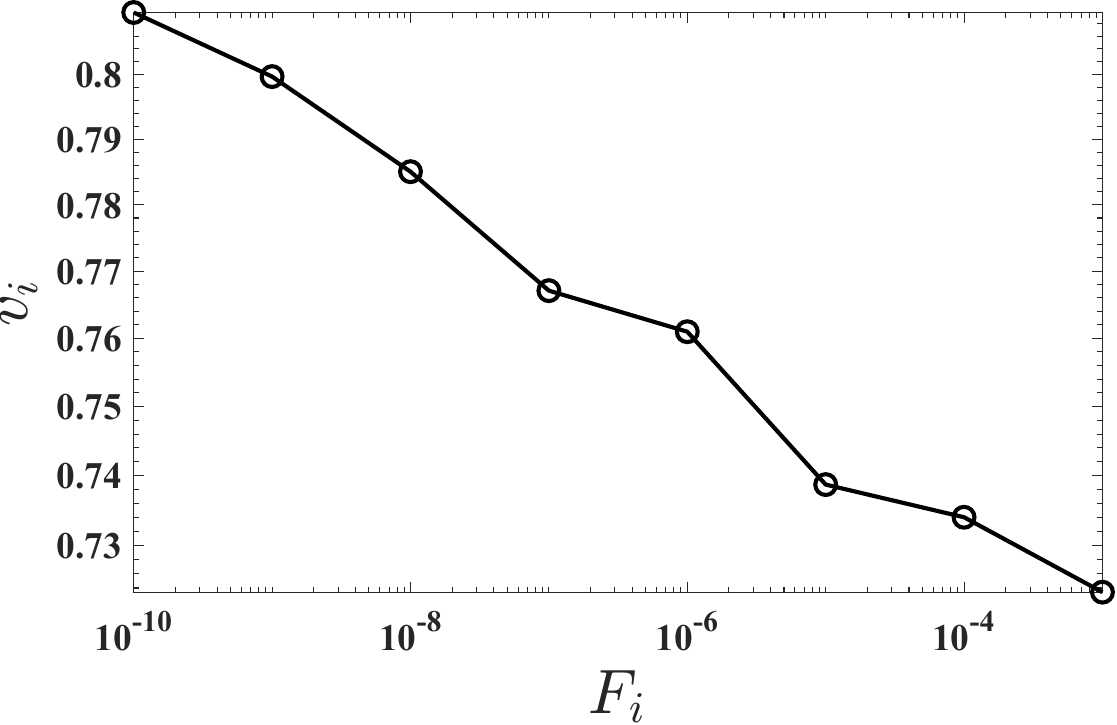}
\caption{The relationship between the invasion velocity and the magnitude of the initial perturbation.}\label{changeInitialValue}
\end{figure}

As an illustrative example, we show the behavior of the invasion phenomenon in the supercritical region in Fig.~\ref{phaseSpinvasionSuper}. Here we choose the spatial scale $L_x=600$ to demonstrate that, in the supercritical region, the invasion phenomenon also exhibits a cutoff behavior when the spatial scale is sufficiently large. The cutoff phenomenon here may originate from the stochastic fluctuations of the system. During the evolution of the system, these fluctuations are amplified by the phase separation effect, leading to the cessation of the invasion phenomenon beyond a certain time scale. For comparison, we also consider a configuration where the system is not divided in half but is partitioned as shown in the middle panel of Fig.~\ref{phaseSpinvasionSuper}. The results show that the division method in the $x$ direction does not affect the invasion phenomenon. Meanwhile, we also show in Fig.~\ref{changeInitialValue} the evolution of $\psi(x)$ under different quench parameters. As the system approaches the critical point, the invasion velocity decreases, and correspondingly, the distance between adjacent bubbles increases. Owing to the larger distance between bubbles, the maximum condensate $\max(\langle |\mathcal{O}| \rangle)$ becomes larger, because the normal phase occupies a larger proportion and thus effectively gains more condensate. In addition, we investigate the effect of different initial perturbation amplitudes $F_i$ on the invasion velocity, and present the results in Fig.~\ref{changeInitialValue}.




\begin{figure}
\centering
\includegraphics[width=1\columnwidth]{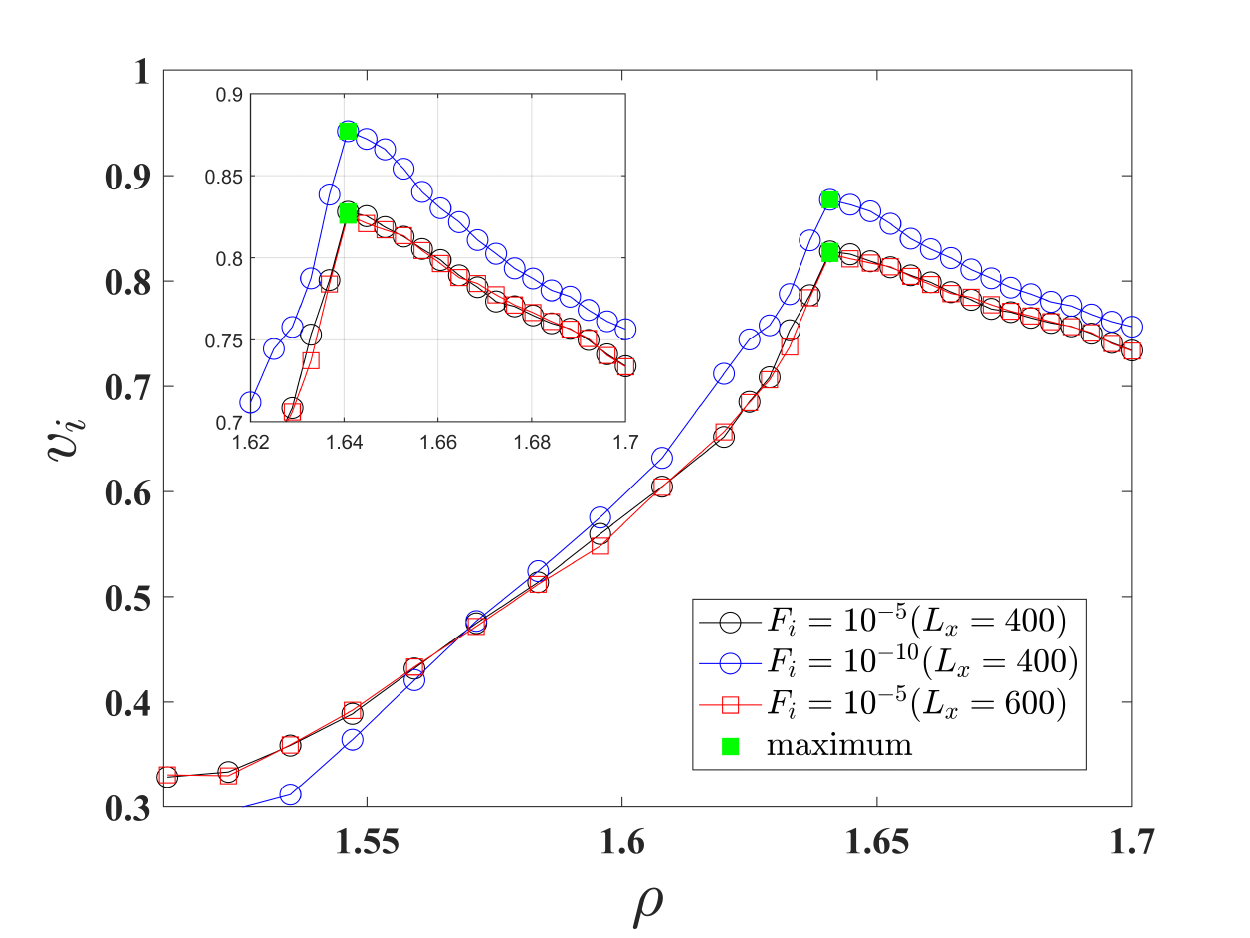}
\caption{The relationship between the invasion velocity in the supercritical region and the charge density $\rho$ for $\tau=2.93$.}\label{exampleTau2o93}
\end{figure}

\begin{figure}
\centering
\includegraphics[width=1\columnwidth]{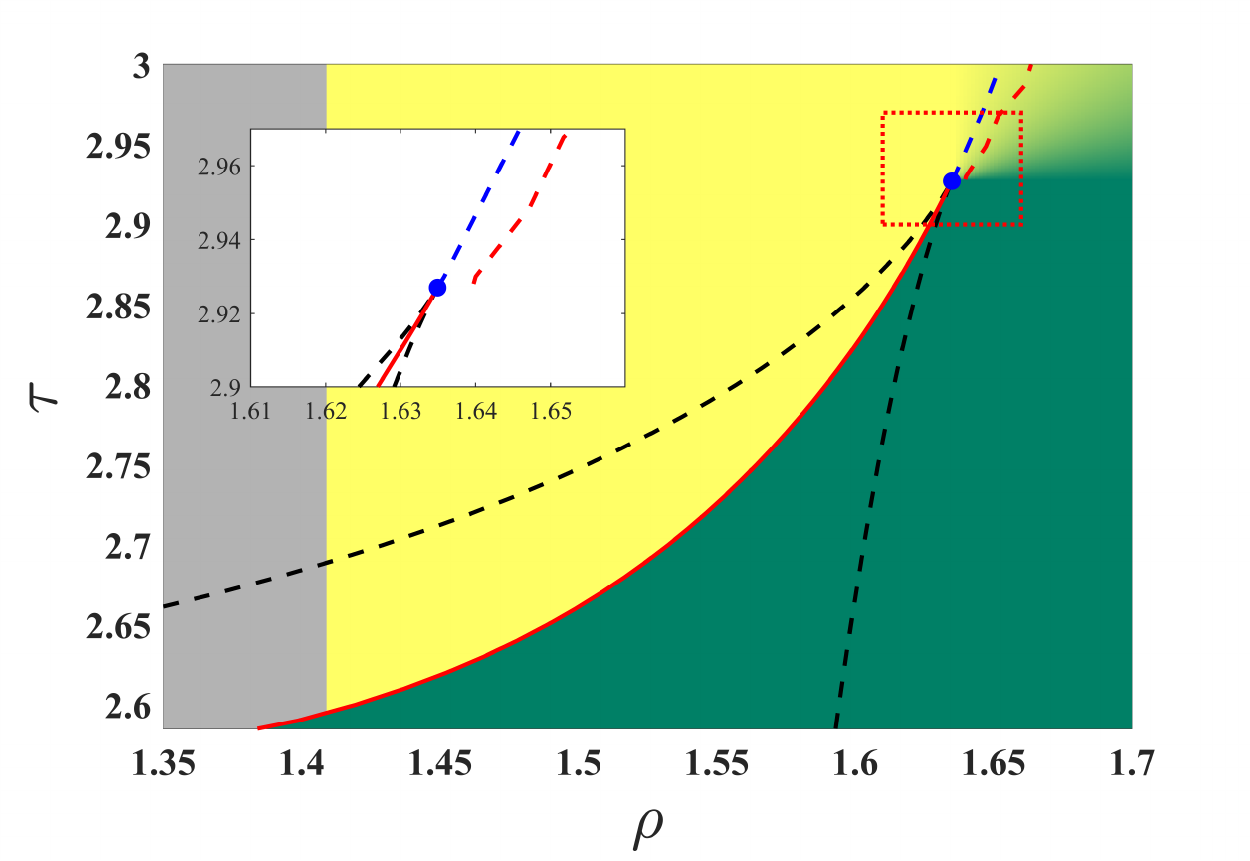}
\caption{The phase diagram of holographic system in the $\tau$-$\rho$ parameter space. The red solid line denotes the first-order phase transition points, the black dashed line denotes the spinodal region of the first-order phase transition, the blue point denotes the critical point, the blue dashed line denotes the supercritical crossover curve defined by the pure phase separation effect, and the red dashed line denotes the crossover line defined by the invasion velocity in the supercritical region. The gray region represents the normal solution, while the yellow and green regions represent two different condensate solutions.}\label{supercriticalDiagram}
\end{figure}


Subsequently, we computed the relationship between the invasion velocity and the quench endpoint $\rho_f$ for a fixed $\tau = 2.93$, and the results are presented in Fig.~\ref{exampleTau2o93}. The initial quench point was uniformly set to $\rho_s = 1.3$, and the quench rate was fixed at $\tau_Q = 0.1$. The black circles in the figure denote the numerically obtained values. As can be observed, the invasion velocity exhibits a distinct inflection point within the supercritical region, indicated by the green square in Fig.~\ref{exampleTau2o93}. It can be seen that below the green square, the invasion velocity increases with $\rho_f$, reaches a maximum at the green square, and then decreases as $\rho_f$ further increases. This indicates that in the supercritical region, the invasion velocity can distinguish two different supercritical subphases by its variation trend. In addition, we also compute the results of $v_i$ versus $\rho_f$ in the supercritical region for the initial perturbation $F_i = 10^{-10}$ and system size $L_x = 600$. These results are shown as blue circles and red squares in Fig.~\ref{exampleTau2o93}. It can be seen that although there is some quantitative change, the magnitude of the initial perturbation does not affect the position of $\rho_f$ at which the maximum invasion velocity occurs in the supercritical region.


\subsection{Phase diagram}
Finally, we compute the time for the emergence of inhomogeneous structures and the invasion velocity for different values of $\tau$, as well as their dependence on the quench endpoint $\rho_f$. We then identify the dynamical crossover points for these parameters. These points are marked by the blue and red dashed lines in Fig.~\ref{supercriticalDiagram}. When the system parameters exceed the critical point indicated by the blue dot, the system enters the supercritical region where the two superfluid phases become indistinguishable, and the first-order phase transition can no longer distinguish the two different supercritical region. However, by computing the time for the most rapid emergence of inhomogeneous structures and the maximum invasion velocity in the supercritical region, we can clearly define two supercritical crossover curves that contain mixed thermodynamic and dynamical information. These two curves differ from the traditional Widom line or Frenkel line, as they simultaneously incorporate nonequilibrium information from both thermodynamics and dynamics, which allows us to study the physics of the supercritical region from a completely new perspective.

It is worth noting that the supercritical crossover line defined by the pure phase separation effect passes through the critical point, whereas the line defined by the invasion velocity does not. The invasion velocity contains mixed information from both symmetry breaking and phase separation, making the underlying physical process inherently complex. Nevertheless, the fact that it does not pass through the critical point does not conflict with existing theories, since conventionally used dynamically defined crossover lines, such as the Frenkel line \cite{Yoon_2018,PhysRevLett.111.145901,Bolmatov2013,Prescher_2017,Bolmatov_2015,Fomin_2018,Fomin2015,PhysRevE.85.031203,2023PhRvR...5a3149H,jiang2024experimental}, also do not pass through the critical point. Moreover, neither of these two curves is inherently superior to the other, as they account for different physical processes.


\section{Discussion}\label{sec5}
From the perspective of nonequilibrium dynamics, this work proposes a novel approach to characterize the supercritical region. Traditional studies distinguishing supercritical subphases mainly rely on static thermodynamic response functions (such as specific heat and compressibility) or equilibrium correlation functions, which are essentially based on quasi-static processes. In contrast, we focus on the dynamical evolution after a rapid quench of the system. We find that both the pure phase separation effect and the uniform invasion phenomenon induced by topological defects still exist in the supercritical region. Moreover, the time for phase separation to generate inhomogeneous structures and the invasion velocity exhibit a clear turning point as functions of the quench endpoint $\rho_f$. These two turning points define two new curves in the parameter space, which we call the nonequilibrium supercritical crossover lines.

Compared with the classical supercritical crossover lines, such as the Widom line or the Frenkel line, the crossover lines proposed in this work differ fundamentally. They are not defined by extrema of static response functions or by quasistatic relaxation behaviors. Instead, they originate from the phase separation dynamics of the system or from the coupling effect with internal symmetry breaking, and contain both thermodynamic information and dynamical information. Therefore, these lines reflect the dynamical effects of the supercritical region under nonequilibrium conditions.


Importantly, the phenomena studied here do not depend on the specific details of the holographic model. The underlying physical mechanism can be attributed to symmetry breaking, phase separation, and quench dynamics. For systems without symmetry breaking, the supercritical crossover point can be identified by the time at which inhomogeneous structures appear most rapidly after the quench. For systems that involve both symmetry breaking and phase separation, the invasion velocity, which simultaneously contains the effects of both processes, can be used to determine the location of the supercritical crossover. Notably, recent nonequilibrium experiments by Lee et al. \cite{Lee2025} on supercritical fluids observed analogous long-lived cluster structures, providing experimental support for the supercritical phase separation mechanism. Moreover, the invasion effect does not necessarily require topological defects to be triggered. Refs.~\cite{PhysRevE.48.2861,Scheel2017} demonstrate that interface boundaries can also induce invasion. Therefore, for systems without symmetry breaking, it is in principle still possible to probe the properties of the supercritical region via the invasion velocity.


In summary, from the perspective of nonequilibrium dynamics, we have defined two new crossover lines in the supercritical region for two different scenarios. This reveals that even in the supercritical region where no phase transition occurs, the dynamical evolution can still exhibit distinct subphases. This finding not only broadens our understanding of supercritical subphases but also provides an actionable bridge between theoretical models and experimental observations. Future work may verify this result through numerical simulations in other systems, test the universality of this crossover line across diverse physical systems, and carry out corresponding experiments to validate the predictions made here.

\section*{Acknowledgement}
This work is supported by the National Natural Science Foundation of China (grant nos. 12575054, 12533001, 12575049, 12473001, 12205039, 12305058, and 11965013). ZYN is partially supported by Yunnan High-level Talent Training Support Plan Young $\&$ Elite Talents Project (grant no. YNWR-QNBJ-2018-181). This work is also supported by the National SKA Program of China (grant nos. 2022SKA0110200 and 2022SKA0110203) and the 111 Project (grant no. B16009).

\bibliographystyle{apsrev4-1}
\bibliography{NCSqd}





\end{document}